\documentclass[12pt]{article}
\usepackage{epsfig}
\def\be{\begin{equation}}
\def\ee{\end{equation}}
\def\bea{\begin{eqnarray}}
\def\eea{\end{eqnarray}}
\usepackage{graphicx}

\catcode`\@=11
\def\lsim{\mathrel{\mathpalette\@versim<}}
\def\gsim{\mathrel{\mathpalette\@versim>}}
\def\@versim#1#2{\vcenter{\offinterlineskip
\ialign{$\m@th#1\hfil##\hfil$\crcr#2\crcr\sim\crcr } }}
\catcode`\@=12

\catcode`@=12
\begin{document}
\thispagestyle{empty}
\begin{flushright}
UCRHEP-T486\\
February 2010\
\end{flushright}
\vspace{0.3in}
\begin{center}
{\LARGE \bf Bound on $Z'$ Mass from CDMS II\\
in the Dark Left-Right Gauge Model II\\}
\vspace{0.8in}
{\bf Shaaban Khalil$^{a,b}$, Hye-Sung Lee$^c$, and Ernest Ma$^d$\\}
\vspace{0.2in}
{\sl $^a$ Centre for Theoretical Physics, The British University in Egypt,\\
El Sherouk City, Postal No.~11837, P.O.~Box 43, Egypt\\}
\vspace{0.1in}
{\sl $^b$ Department of Mathematics, Ain Shams University,\\
Faculty of Science, Cairo 11566, Egypt\\}
\vspace{0.1in}
{\sl $^c$ Department of Physics, Brookhaven National Laboratory, 
Upton, New York 11973, USA\\}
\vspace{0.1in}
{\sl $^d$ Department of Physics and Astronomy, University of California,\\
Riverside, California 92521, USA\\}
\end{center}
\vspace{0.8in}
\begin{abstract}\
With the recent possible signal of dark matter from the CDMS II experiment, 
the $Z'$ mass of a new version of the dark left-right gauge model (DLRM II) 
is predicted to be at around a TeV.  As such, it has an excellent discovery 
prognosis at the operating Large Hadron Collider.
\end{abstract}

\newpage
\baselineskip 24pt

\noindent \underline{\it Introduction}~:~ One year ago, we 
proposed~\cite{klm09} that dark-matter fermions (scotinos) are naturally 
present in an unconventional left-right gauge extension of the standard model 
(SM) of particle interactions, which we call the dark left-right model (DLRM). 
It is a nonsupersymmeric variation of the alternative left-right model (ALRM) 
discussed already 23 years ago~\cite{m87,bhm87}.  One important difference 
of both the DLRM and the ALRM with the conventional left-right model 
(LRM)~\cite{dgko91} is the fact that tree-level flavor-changing neutral 
currents~\cite{gw77} are naturally absent so that the $SU(2)_R$ breaking 
scale may easily be at around a TeV, allowing both the charged $W_R^\pm$ 
and the extra neutral $Z'$ gauge bosons to be observable at the large 
hadron collider (LHC).  Interesting phenomenology of $Z'$ decay into 
scalar bosons in the DLRM has just recently been discussed~\cite{adhm10}.

In this paper, we propose a new variant of this extension which we call 
DLRM II.  [Other more exotic variants are also possible~\cite{m09}.] 
Instead of having Majorana scotinos as dark matter, we now have 
Dirac scotinos.  Their interactions with nuclei through the $Z'$ are thus 
relevant for understanding the recent result of the dark-matter direct-search 
experiment CDMS II~\cite{cdms09}.  It will be shown that the $Z'$ mass may 
indeed be around a TeV, and its discovery prognosis at the LHC is excellent.

\noindent \underline{\it Model}~:~ Consider the gauge group $SU(3)_C \times
SU(2)_L \times SU(2)_R \times U(1)$.  The conventional leptonic assignments
are $\psi_L = (\nu,e)_L \sim (1,2,1,-1/2)$ and  $\psi_R = (\nu,e)_R \sim
(1,1,2,-1/2)$.  Hence $\nu$ and $e$ obtain Dirac masses through the
Yukawa terms $\overline{\psi}_L \Phi \psi_R$ and $\overline{\psi}_L
\tilde{\Phi} \psi_R$, where $\Phi = (\phi_1^0, \phi_1^-; \phi_2^+,
\phi_2^0) \sim (1,2,2,0)$ is a Higgs bidoublet and $\tilde{\Phi} =
\sigma_2 \Phi^* \sigma_2 = (\overline{\phi_2^0},-\phi_2^-;-\phi_1^+,
\overline{\phi_1^0})$ transforms in the same way.  Both $\langle \phi_1^0
\rangle$ and $\langle \phi_2^0 \rangle$ contribute to $m_\nu$ and $m_e$,
and similarly $m_u$ and $m_d$ in the quark sector, resulting thus in the
appearance of tree-level flavor-changing neutral currents.

Suppose the term $\overline{\psi}_L \tilde{\Phi} \psi_R$ is forbidden by a
symmetry, then the same symmetry may be used to maintain $\langle \phi_1^0
\rangle = 0$ and only $e$ gets a mass through $\langle \phi_2^0 \rangle
\neq 0$.  At the same time, $\nu_L$ and $\nu_R$ are not Dirac mass partners,
so they could in fact be completely different
particles with independent masses of their own.  Whereas $\nu_L$ is clearly
the neutrino we observe in the usual weak interactions, $\nu_R$ can now be
something else entirely.  Here we rename $\nu_R$ as $n_R$ and show that
it may in fact be a scotino, i.e. a fermionic dark-matter candidate.

In our previous proposal~\cite{klm09}, we imposed a new global $U(1)$ symmetry 
$S$ in such a way that the spontaneous breaking of $SU(2)_R \times S$ will 
leave the combination $L = S - T_{3R}$ unbroken.  We then showed that $L$ is 
a generalized lepton number, with $L=1$ for the known leptons, and $L=0$ 
for all known particles which are not leptons.  Here we consider instead 
the case $L = S + T_{3R}$.  Our model is nonsupersymmetric, but it may be 
rendered supersymmetric by the usual procedure which takes the SM to the MSSM
(minimal supersymmetric standard model).  Under $SU(3)_C \times SU(2)_L
\times SU(2)_R \times U(1) \times S$, the fermions transform as shown
in Table 1. Note the necessary appearance of the exotic quark $h$, which
will turn out to carry lepton number as well.

\begin{table}[htb]
\caption{Fermion content of proposed model.}
\begin{center}
\begin{tabular}{|c|c|c|}
\hline
Fermion & $SU(3)_C \times SU(2)_L \times SU(2)_R \times U(1)$ & $S$ \\
\hline
$\psi_L = (\nu,e)_L$ & $(1,2,1,-1/2)$ & $1$ \\
$\psi_R = (n,e)_R$ & $(1,1,2,-1/2)$ & $3/2$ \\
$\nu_R$ & $(1,1,1,0)$ & $1$ \\ 
$n_L$ & $(1,1,1,0)$ & $2$ \\ 
\hline
$Q_L = (u,d)_L$ & $(3,2,1,1/6)$ & $0$ \\
$Q_R = (u,h)_R$ & $(3,1,2,1/6)$ & $-1/2$ \\
$d_R$ & $(3,1,1,-1/3)$ & $0$ \\
$h_L$ & $(3,1,1,-1/3)$ & $-1$ \\
\hline
\end{tabular}
\end{center}
\end{table}

The scalar sector consists of one bidoublet and two doublets:
\begin{equation}
\Phi = \pmatrix{\phi_1^0 & \phi_2^+ \cr \phi_1^- & \phi_2^0}, ~~~
\Phi_L = \pmatrix{\phi_L^+ \cr \phi_L^0}, ~~~ \Phi_R = \pmatrix{\phi_R^+
\cr \phi_R^0}.
\end{equation}
Their assignments under $S$ are listed in Table 2.

\begin{table}[htb]
\caption{Scalar content of proposed model.}
\begin{center}
\begin{tabular}{|c|c|c|}
\hline
Scalar & $SU(3)_C \times SU(2)_L \times SU(2)_R \times U(1)$ & $S$ \\
\hline
$\Phi$ & $(1,2,2,0)$ & $-1/2$ \\
$\tilde{\Phi} = \sigma_2 \Phi^* \sigma_2$ & $(1,2,2,0)$ & $1/2$ \\
$\Phi_L$ & $(1,2,1,1/2)$ & $0$ \\
$\Phi_R$ & $(1,1,2,1/2)$ & $1/2$ \\
\hline
\end{tabular}
\end{center}
\end{table}

The Yukawa terms allowed by $S$ are then $\overline{\psi}_L \Phi \psi_R$,
$\overline{\psi}_L \tilde{\Phi}_L \nu_R$, $\overline{\psi}_R \tilde{\Phi}_R n_L$,
$\overline{Q}_L \tilde{\Phi} Q_R$, $\overline{Q}_L \Phi_L d_R$, and 
$\overline{Q}_R \Phi_R h_L$,  whereas
$\overline{\psi}_L \tilde{\Phi} \psi_R$, $\overline{n}_L \nu_R$, 
$\overline{Q}_L \Phi Q_R$, and
$\overline{h}_L d_R$ are forbidden.  Hence $m_e$, $m_u$ come from $v_2 =
\langle \phi_2^0 \rangle$; $m_\nu$, $m_d$ come from $v_3 = \langle \phi_L^0 
\rangle$; and $m_n$, $m_h$ come from $v_4 = \langle \phi_R^0 \rangle$.  
This structure shows clearly that flavor-changing neutral currents
are guaranteed to be absent at tree level.

As it stands, both the neutrino $\nu$ and the scotino $n$ are Dirac fermions, 
and lepton number $L$ is conserved.  If we now introduce a soft term 
$\nu_R \nu_R$ which breaks $L$ by two units, then $\nu_L$ gets a Majorana 
mass through the canonical seesaw mechanism, as is usually assumed.
As for $n$, it remains a Dirac fermion, being protected by a residual 
global $U(1)$ symmetry, under which $n$, $W_R^+$ transform as 1, and $h$, 
$\phi_1^{0,-}$ transform as $-1$.

\noindent \underline{\it Gauge sector}~:~ Since $e$ has $L=1$ and $n$ has
$L=2$, the $W_R^+$ of this model must have $L = S + T_{3R} = 0 + 1 = 1$.
This also means that unlike the conventional LRM, $W_R^\pm$ does not mix
with the $W_L^\pm$ of the SM at all.  This important property allows the 
$SU(2)_R$ breaking scale to be much lower than it would be otherwise, 
as explained already 23 years ago \cite{m87,bhm87}.  Let $e/g_L = s_L = \sin 
\theta_W$ and $s_R = e/g_R$, with $c_{L,R} = \sqrt{1-s_{L,R}^2}$, then 
$g_B = e/\sqrt{c_L^2-s_R^2}$ and the 
neutral gauge bosons of the DLRM (as well as the ALRM) are given by
\begin{equation}
\pmatrix{A \cr Z \cr Z'} = \pmatrix{s_L & s_R & \sqrt{c_L^2-s_R^2} \cr
c_L & -s_L s_R/c_L & -s_L \sqrt{c_L^2 - s_R^2}/c_L \cr 0 &
\sqrt{c_L^2 - s_R^2}/c_L & -s_R/c_L} \pmatrix{W_L^0 \cr W_R^0 \cr B}.
\end{equation}
Whereas $Z$ couples to the current $J_{3L} - s_L^2 J_{em}$ with coupling
$e/s_L c_L$ as in the SM, $Z'$ couples to the current
\begin{equation}
J_{Z'} = s_R^2 J_{3L} + c_L^2 J_{3R} - s_R^2 J_{em}
\end{equation}
with coupling $g_{Z'} = e/s_R c_L \sqrt{c_L^2-s_R^2}$.

The masses of the gauge bosons are given by
\begin{eqnarray}
&& M_{W_L}^2 = {e^2 \over 2 s_L^2} (v_2^2 + v_3^2), ~~~ M_Z^2 = {M_{W_L}^2 \over 
c_L^2},
~~~ M_{W_R}^2 = {e^2 \over 2 s_R^2} (v_4^2 + v_2^2), \\
&& M_{Z'}^2 = {e^2 c_L^2 \over 2 s_R^2  (c_L^2 - s_R^2)} (v_4^2 + v_2^2) -
{s_L^2 s_R^2 M_{W_L}^2 \over c_L^2(c_L^2 - s_R^2)},
\end{eqnarray}
where zero $Z-Z'$ mixing has been assumed, using the condition
\cite{bhm87} $v_2^2/(v_2^2+v_3^2) = s_R^2/c_L^2$.

\noindent \underline{\it Direct search constraint from CDMS II}~:~
The $Z'$ couplings to $u$, $d$, $n$ (in units of $g_{Z'}$) are given by
\begin{eqnarray}
&& u_L = -{1 \over 6} s_R^2, ~~~ u_R = {1 \over 2} c_L^2 - {2 \over 3} s_R^2, 
~~~ u_V = {1 \over 4} c_L^2 - {5 \over 12} s_R^2, \\
&& d_L = -{1 \over 6} s_R^2, ~~~ d_R = {1 \over 3} s_R^2, ~~~ d_V = 
{1 \over 12} s_R^2, \\ 
&& n_L = 0, ~~~ n_R = {1 \over 2} c_L^2, ~~~ n_V = {1 \over 4} c_L^2.
\end{eqnarray}
The effective Lagrangian for elastic scattering of the scotino $n$ off 
quarks is then given by
\begin{equation}
{\cal L} = {g^2_{Z'} n_V \over M^2_{Z'}} (\bar{n} \gamma_\mu n)(u_V \bar{u} 
\gamma^\mu u + d_V \bar{d} \gamma^\mu d).
\end{equation}
In the original DLRM~\cite{klm09}, $n$ is a Majorana scotino, so it does not 
contribute to the s-wave elastic spin-independent scattering cross section 
in the nonrelativistic limit.  Here $n$ is a Dirac scotino, so it will 
contribute. Let
\begin{equation}
f_P = g^2_{Z'} n_V (2 u_V + d_V)/M^2_{Z'}, ~~~ 
f_N = g^2_{Z'} n_V (u_V + 2 d_V)/M^2_{Z'},
\end{equation}
then its elastic cross section per nucleon is given by~\cite{cls09}
\begin{equation}
\sigma_0 = {4 m_r^2 \over \pi} {[Z f_P + (A - Z) f_N]^2 \over A^2},
\end{equation}
where $Z$ and $A$ are the atomic and mass numbers of the target nucleus, 
and $m_r = m_n m_P/(m_n + m_P) \simeq m_P$.  The CDMS II 
collaboration~\cite{cdms09} observed two possible signal events with an 
expected background of $0.6 \pm 0.1$.  Using $^{73}$Ge, i.e. $Z = 32$ 
and $A - Z = 41$, as a representative estimate of $\sigma_0$, this result 
could also be considered as an upper bound, i.e.
\begin{equation}
\sigma_0 = {\pi \alpha^2 m_P^2 (105 c_L^2 - 137 s_R^2)^2 \over (146)^2 
s_R^4 (c_L^2 - s_R^2)^2 M^4_{Z'}} < 3.8 \times 10^{-8}~{\rm pb},
\end{equation}
which occurs at $m_n = 70$ GeV.

\noindent \underline{\it Phenomenological analysis}~:~  We consider the range 
$e^2 < s_R^2 < c_L^2 - e^2$, where the lower bound corresponds to $g_R=1$ 
and the upper bound to $g_B=1$.  The values of $g_{Z'}$ and $\Gamma_{Z'}/M_{Z'}$ 
are plotted in Fig.~1(a) and (b), where $Z'$ is assumed to decay only into 
SM fermions.

\begin{figure}[htb]
\begin{center}
\includegraphics[width=0.45\textwidth]{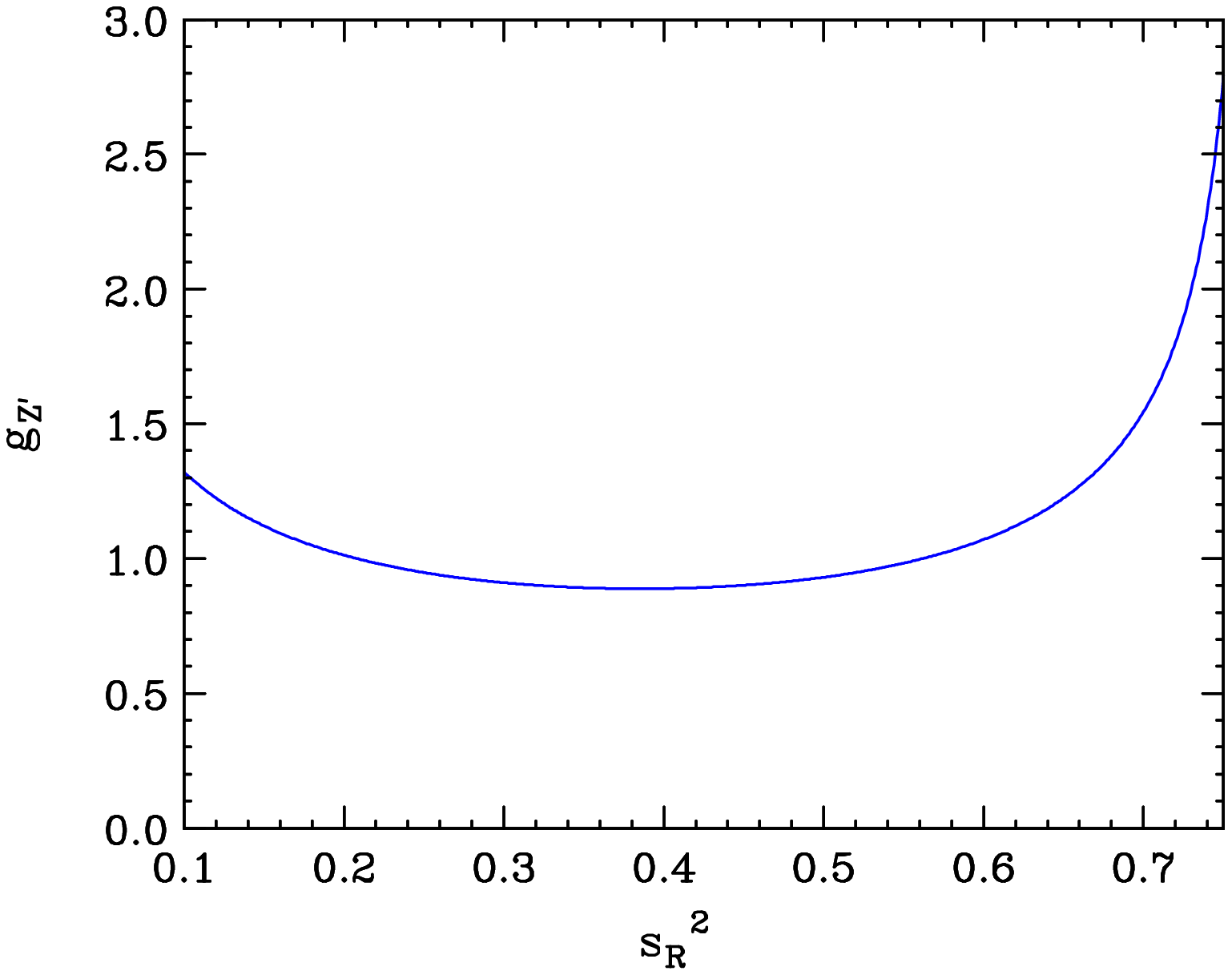} ~~~~~~
\includegraphics[width=0.45\textwidth]{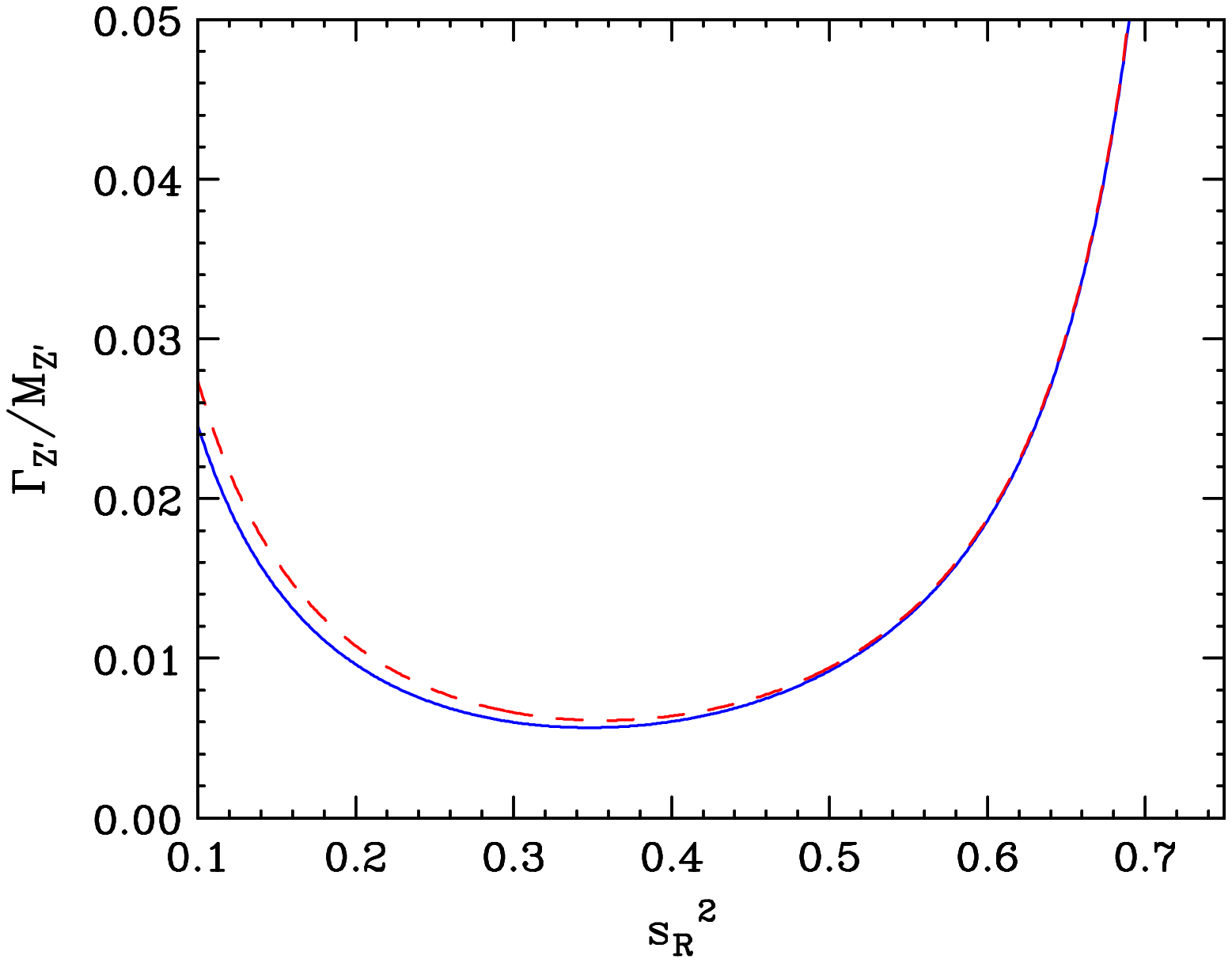} \\
(a) ~~~~~~~~~~~~~~~~~~~~~~~~~~~~~~~~~~~~~~~~~~~~~~~~~~~~~~~~~~~~~~~~~~~~~~~~~~~~~~~~~ (b)
\end{center}
\caption{(a) $g_{Z'}$ vs $s_R^2$. (b) $\Gamma_{Z'} / M_{Z'}$ vs $s_R^2$ for SM 
fermions decay products only in the cases $M_{Z'} = 500$ GeV (blue solid) and $M_{Z'} \to \infty$ 
(red dashed).}
\label{fig:plot1}
\end{figure}

We compute the production and decay of $Z'$ to $e^+e^-$ at the Tevatron as 
a function of $M_{Z'}$ for various values of $s_R^2$ and compare it to 
data~\cite{Aaltonen:2008vx} at $E_{\rm cm} = 1.96$ TeV and an integrated 
luminosity of 2.5 fb$^{-1}$ in Fig.~2(a). We then plot the exclusion limits 
on $M_{Z'}$ from both the new CDMS II data and the Tevatron as a function 
of $s_R^2$ in Fig.~2(b).  Note that the CDMS II bound is stronger than the 
Tevatron bound for $s_R^2 < 0.5$.  Note also that due to the accidental 
cancellation in the numerator of $\sigma_0$ in Eq.~(12), the observed events 
at CDMS II cannot be interpreted as signals of dark matter in this model 
if $s_R^2 > 0.5$, because they would be excluded by the Tevatron data.

\begin{figure}[htb]
\begin{center}
\includegraphics[width=0.45\textwidth]{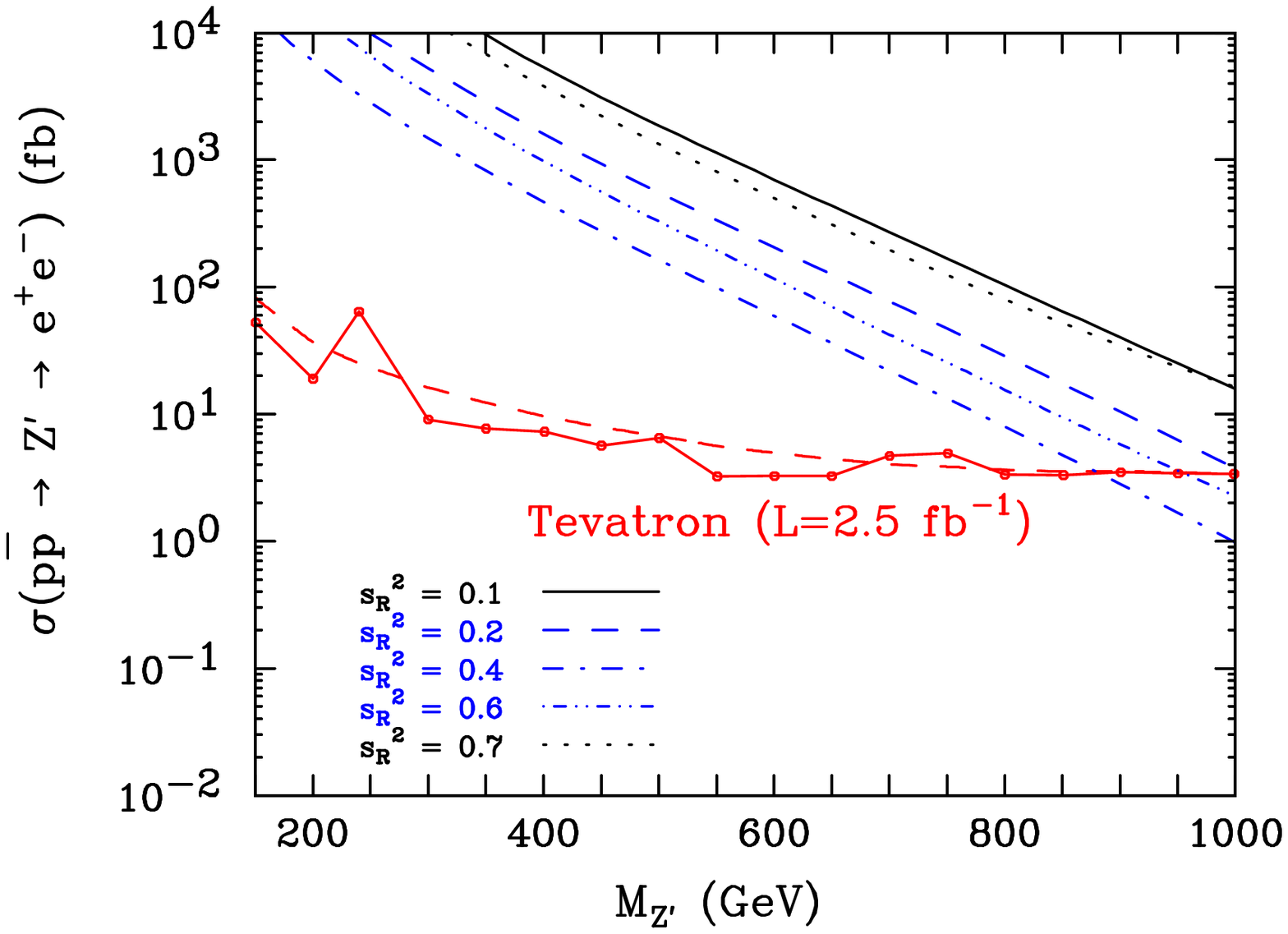} ~~~~~~
\includegraphics[width=0.45\textwidth]{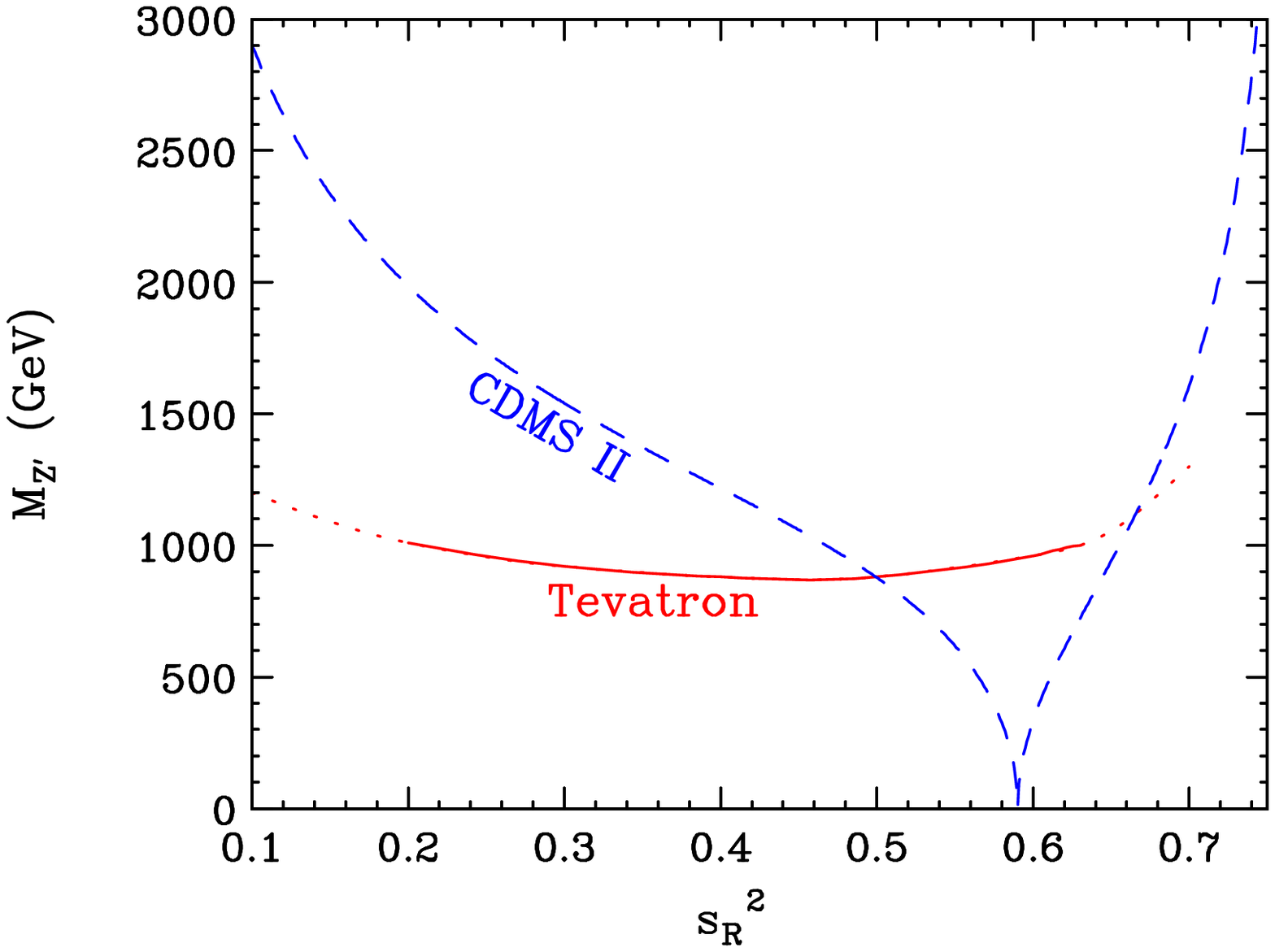}\\
(a) ~~~~~~~~~~~~~~~~~~~~~~~~~~~~~~~~~~~~~~~~~~~~~~~~~~~~~~~~~~~~~~~~~~~~~~~~~~~~~~~~~ (b)
\end{center}
\caption{(a) Lower bound on the $Z'$ mass in this model from Tevatron 
dielectron search. (b) $M_{Z'}$ vs $s_R^2$ from the CDMS II (blue dashed) and 
Tevatron (red solid) bounds. The dotted segments assume a simple extrapolation 
of the Tevatron data.} 
\label{fig:collider}
\end{figure}

Given that $M_{Z'}$ is allowed to be in the TeV range, its discovery 
prognosis is excellent at the LHC.  We show in Fig.~3 its discovery reach 
(assuming $E_{\rm cm} = 14$ TeV) by 10 dilepton events (either dielectron 
or dimuon) which satisfy the following basic cuts on their transverse momenta, 
rapidities, and invariant mass: 
$p_T > 20$ GeV (each lepton), 
$|\eta| < 2.4$ (each lepton), 
$|M_{\ell {\bar \ell}} - M_{Z'}| < 3 \Gamma_{Z'}$.

Using these cuts, the dominant SM background from $\gamma/Z$ (Drell-Yan) 
is negligible.  With an integrated luminosity of 1 fb$^{-1}$, the 
$Z'$ of DLRM II with $M_{Z'} \sim 2$ TeV may then be discovered at the LHC.
\begin{figure}[htb]
\begin{center}
\includegraphics[width=0.55\textwidth]{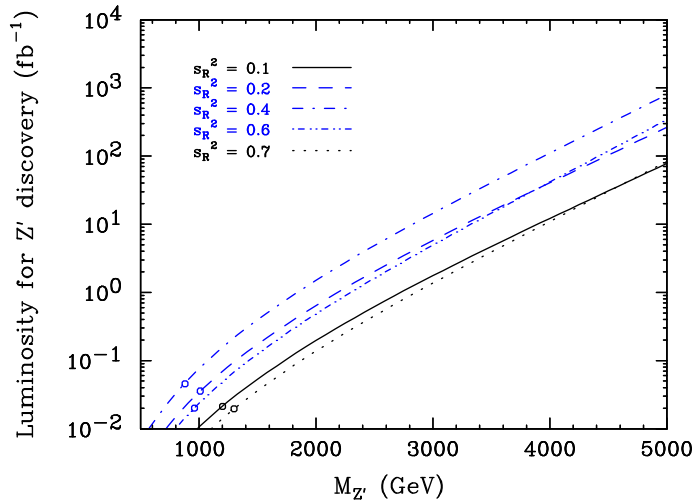}
\end{center}
\caption{Luminosity for $Z'$ discovery by 10 dielectron events at LHC. Small circles are Tevatron limits.}
\label{fig:LHC}
\end{figure}

\noindent \underline{\it Dark-matter relic abundance}~:~ In this model, the 
dark-matter relic abundance is presumably determined by the annihilation 
$n \bar{n} \to Z' \to$ SM fermions.  The thermally averaged cross section 
multiplied by relative velocity is approximately given by
\begin{equation}
\langle \sigma v_{rel} \rangle_{Z'} = {g^4_{Z'} c_L^4 m_n^2 \sum_f (f_L^2 + f_R^2) 
\over 32 \pi (4 m_n^2 - M_{Z'}^2)^2},
\end{equation}
where the sum over fermions should include a factor of 3 for quarks and 
an overall factor of 3 for families.
Fixing the above at 1 pb as a typical value to satisfy the requirement 
of dark-matter relic abundance, it can easily be shown that for $m_n=70$ 
GeV, the required $M_{Z'}$ is very much below the CDMS II bound. [For example, 
for $s_R^2=0.4$, $M_{Z'} = 267$ GeV would be required.]  In other words, 
the $n \bar{n} \to Z'$ annihilation cross section would be too small to 
account for the observed dark-matter relic abundance.  To remedy this 
situation, the mechanism proposed in the original DLRM may be invoked, 
i.e. $n \bar{n} \to l^- l^+$ through $\Delta_R^+$ exchange.  However, 
this requires adding the $SU(2)_R$ scalar triplet $(\Delta_R^{++},\Delta_R^+,
\Delta_R^0)$, which is not necessary in our present version and thus not 
very much motivated.  The alternative is to consider a larger value of 
$m_n$.

The CDMS II bound on $\sigma_0$ is very well approximated in the range 
$0.3 < m_n < 1.0$ TeV by the expression
\begin{equation}
\sigma_0 < 2.2 \times 10^{-7}~{\rm pb}~(m_n/1~{\rm TeV})^{0.86}.
\end{equation}
Using this on the right-hand side of Eq.~(12), we plot in Fig.~4(a) and 
(b) the $M_{Z'}$ bounds for $m_n=400$ and 600 GeV, as well as the solutions 
of $M_{Z'}$ (with $M_{Z'} > 2m_n$) to Eq.~(13) for 1 pb.  We see that 
there are indeed consistent solutions (where the solid curve is higher 
than the dash curve) for a range of $s_R^2$ in each case.  If $m_n$ 
falls below 300 GeV, then there is no solution because $M_{Z'}$ would 
then be excluded by the Tevatron bound.  We note also that 
only a modest resonance enhancement is needed from the denominator of 
Eq.~(13).  The $n \bar{n}$ annihilation to $l^+l^-$ through $W_R$ exchange 
also contributes to the dark-matter relic abundance, but its value is 
an order of magnitude less, i.e.
\begin{equation}
\langle \sigma v_{rel} \rangle_{W_R} = {3 g^4_R m_n^2  \over 64 \pi 
(m_n^2 + M_{W_R}^2)^2}.
\end{equation}
\begin{figure}[htb]
\begin{center}
\includegraphics[width=0.45\textwidth]{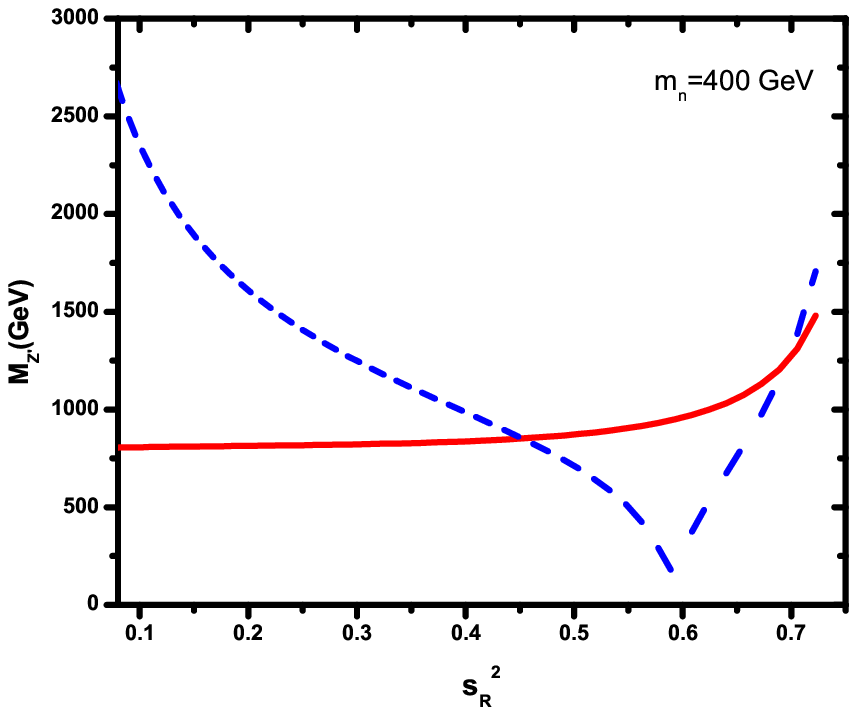} ~~~~~~
\includegraphics[width=0.45\textwidth]{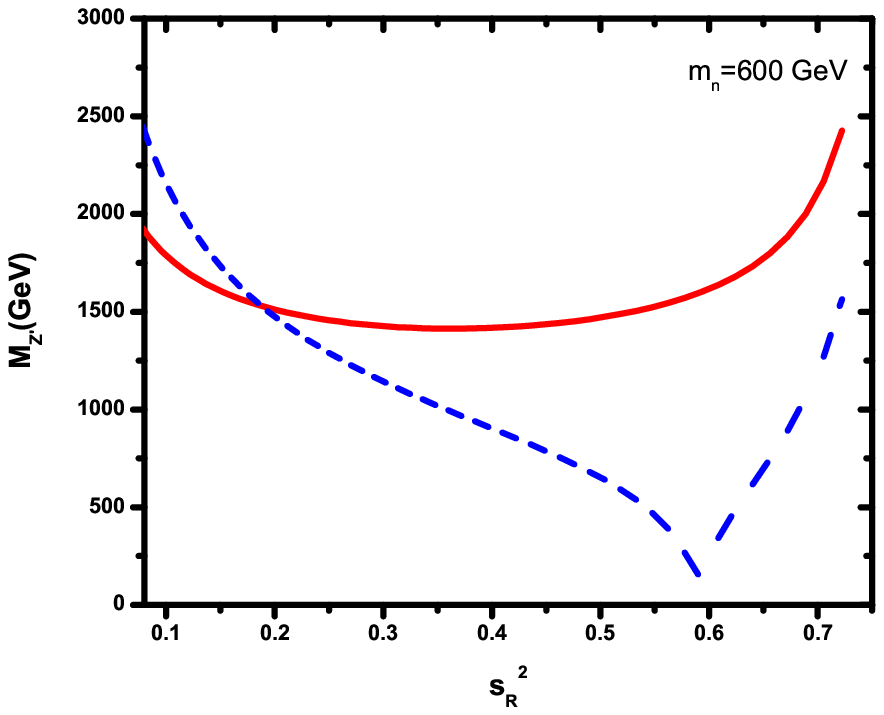} \\
(a) ~~~~~~~~~~~~~~~~~~~~~~~~~~~~~~~~~~~~~~~~~~~~~~~~~~~~~~~~~~~~~~~~~~~~~~~~~~~~~~~~~ (b)
\end{center}
\caption{(a) For $m_n = 400$ GeV, the CDMS II bound on $M_{Z'}$ (blue dashed) 
and the value of $M_{Z'}$ (red) from $\langle \sigma v_{rel} \rangle_{Z'} = 1$ 
pb vs $s_R^2$; (b) same as in (a) for $m_n = 600$ GeV.} 
\label{fig4}
\end{figure}

\noindent \underline{\it Lepton flavor violation}~:~ Unlike the original 
DLRM, where a scalar triplet $(\Delta_R^{++},\Delta_R^+,\Delta_R^0)$ may 
mediate lepton flavor violating processes such as $\mu \to eee$ at tree level, 
and must be forbidden by hand, the DLRM II is safe because it has no such 
interactions. Nevertheless, lepton (as well quark) flavor violation occurs 
in one loop in the $SU(2)_R$ sector, in complete analogy to that of the SM 
in the $SU(2)_L$ sector.  The branching fraction of $\mu \to e \gamma$ is then 
\begin{equation}
B(\mu \to e \gamma) = {3 \alpha |\delta_R|^2 \over 64 \pi} \left( 
{s_L^2 M_{W_L}^2 \over s_R^2 M_{W_R}^2} \right)^2 < 1.2 \times 10^{-11},
\end{equation}
where the experimental upper bound has also been displayed, and $\delta_R$ 
is the analog of the well-known suppression factor $\delta_L = 
\sum_i U^*_{ei} U_{\mu i} (m_{\nu_i}^2/M_{W_L}^2)$ in the SM. 
For $s_R^2 = s_L^2$, we have $M_{W_R} = 1.5$ TeV, then $|\delta_R| < 0.116$. 
Since the flavor structure of scotino mixing and their mass-squared 
differences are unknown, this upper bound could be saturated, and the 
observation of $\mu \to e \gamma$ may be imminent.  The 
same holds for other lepton flavor violating processes such as $\mu - e$ 
conversion in nuclei.  Note that the contribution to the muon anomalous 
magnetic moment here is about $10^{-10}$, well below the 
experimental sensitivity. A more comprehensive study, including contributions 
to $D^0 - \bar{D}^0$ mixing~\cite{m88}, will be given elsewhere.

\noindent \underline{\it Connecting the $Z'$ and dark-matter searches}~:~ 
As the LHC begins its operation, one of its first possible discoveries could 
be a $Z'$ through the process $q \bar q \to Z' \to l^+ l^-$.  There are 
many $Z'$ models, and some of them could also be invoked~\cite{Lee:2008em}
to explain the CDMS II results.  However, the coupling of the dark matter 
to the $Z'$ in these models is in general not related to the $Z'$ 
leptonic couplings.  Here they are intimately connected and predicted 
as a function of only $s_R^2$.  In fact, if we assume $s_R^2 = s_L^2$ 
(i.e. left-right symmetry), then there is no free parameter.  Our numerical 
analysis in this paper is only a rough estimate for illustration, but it 
points to the important assertion that the $Z'$ interactions in this model 
are fixed with respect to direct dark-matter search and the detection of 
$Z'$ itself at an accelerator.  In these exciting times of having both the 
functioning LHC and ongoing dark-matter search experiments, the dark-matter 
mystery in astroparticle physics may be near a solution.

\noindent \underline{\it Acknowledgements}~:~
The work of S.K. was partially supported by the Science and Technology 
Development Fund (STDF) Project ID 437 and the ICTP Project ID 30.
This work was also supported by the U.~S.~Department of Energy Grants 
No. DE-AC02-98CH10886 (HL) and No. DE-FG03-94ER40837 (EM).

\bibliographystyle{unsrt}

\end{document}